\documentclass[twoside,twocolumn,9pt]{article}
\usepackage{extsizes}
\usepackage[super,sort&compress,comma]{natbib} 
\usepackage[version=3]{mhchem}
\usepackage[left=1.5cm, right=1.5cm, top=1.785cm, bottom=2.0cm]{geometry}
\usepackage{balance}
\usepackage{times,mathptmx}
\usepackage{sectsty}
\usepackage{graphicx} 
\usepackage{lastpage}
\usepackage[format=plain,justification=justified,singlelinecheck=false,font={stretch=1.125,small,sf},labelfont=bf,labelsep=space]{caption}
\usepackage{float}
\usepackage{fancyhdr}
\usepackage{fnpos}
\usepackage[english]{babel}
\usepackage{array}
\usepackage{droidsans}
\usepackage{charter}
\usepackage[T1]{fontenc}
\usepackage[usenames,dvipsnames]{xcolor}
\usepackage{setspace}
\usepackage[compact]{titlesec}
\usepackage{hyperref}
\usepackage{endnotes}

\usepackage{epstopdf}

\definecolor{cream}{RGB}{222,217,201}

\begin{document}

\pagestyle{fancy}
\thispagestyle{plain}
\fancypagestyle{plain}{

\fancyhead[C]{\includegraphics[width=18.5cm]{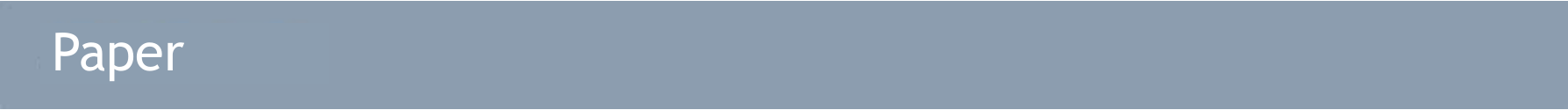}}
\renewcommand{\headrulewidth}{0pt}
}

\makeFNbottom
\makeatletter
\renewcommand\LARGE{\@setfontsize\LARGE{15pt}{17}}
\renewcommand\Large{\@setfontsize\Large{12pt}{14}}
\renewcommand\large{\@setfontsize\large{10pt}{12}}
\renewcommand\footnotesize{\@setfontsize\footnotesize{7pt}{10}}
\makeatother

\renewcommand{\thefootnote}{\fnsymbol{footnote}}
\renewcommand\footnoterule{\vspace*{1pt}%
\color{cream}\hrule width 3.5in height 0.4pt \color{black}\vspace*{5pt}} 
\setcounter{secnumdepth}{5}

\makeatletter 
\renewcommand\@biblabel[1]{#1}            
\renewcommand\@makefntext[1]%
{\noindent\makebox[0pt][r]{\@thefnmark\,}#1}
\makeatother 
\renewcommand{\figurename}{\small{Fig.}~}
\sectionfont{\sffamily\Large}
\subsectionfont{\normalsize}
\subsubsectionfont{\bf}
\setstretch{1.125} 
\setlength{\skip\footins}{0.8cm}
\setlength{\footnotesep}{0.25cm}
\setlength{\jot}{10pt}
\titlespacing*{\section}{0pt}{4pt}{4pt}
\titlespacing*{\subsection}{0pt}{15pt}{1pt}

\fancyfoot{}
\fancyhead{}
\renewcommand{\headrulewidth}{0pt} 
\renewcommand{\footrulewidth}{0pt}
\setlength{\arrayrulewidth}{1pt}
\setlength{\columnsep}{6.5mm}
\setlength\bibsep{1pt}

\makeatletter 
\newlength{\figrulesep} 
\setlength{\figrulesep}{0.5\textfloatsep} 

\newcommand{\topfigrule}{\vspace*{-1pt}%
\noindent{\color{cream}\rule[-\figrulesep]{\columnwidth}{1.5pt}} }

\newcommand{\botfigrule}{\vspace*{-2pt}%
\noindent{\color{cream}\rule[\figrulesep]{\columnwidth}{1.5pt}} }

\newcommand{\dblfigrule}{\vspace*{-1pt}%
\noindent{\color{cream}\rule[-\figrulesep]{\textwidth}{1.5pt}} }

\makeatother

\twocolumn[
  \begin{@twocolumnfalse}
\vspace{3cm}
\sffamily
\begin{tabular}{m{4.5cm} p{13.5cm} }

 & \noindent\LARGE{\textbf{Discontinuous change from thermally- to geometrically - dominated 
effective interactions in colloidal solutions}} \\
\vspace{0.3cm} & \vspace{0.3cm} \\

 & \noindent\large{Nicoletta Gnan,$^{\ast}$\textit{$^{a,b}$} Francesco Sciortino,\textit{$^{a,b}$} and Emanuela Zaccarelli\textit{$^{a,b}$}} \\

& \noindent\normalsize{We report numerical results for the effective potential arising between two colloids immersed in a self-assembling cosolute which forms reversible clusters. The potential is evaluated at  cosolute state points with different densities and temperatures but with the same connectivity properties. We find that the range of the resulting effective potential is controlled only by the cosolute thermal correlation length rather than by its connectivity length.
 We discuss the significant differences with previous results focused on cosolute forming irreversible clusters  and we show that  irreversible bond case represents a singular limit  which cannot be accessed  in equilibrium by continuously increasing the bond lifetime.} \\

\end{tabular}

 \end{@twocolumnfalse} \vspace{0.6cm}

  ]

\renewcommand*\rmdefault{bch}\normalfont\upshape
\rmfamily
\section*{}
\vspace{-1cm}


\footnotetext{\textit{$^{\ast}$ nicoletta.gnan@roma1.infn.it}}
\footnotetext{\textit{$^{a}$~CNR-ISC UOS {\em Sapienza}, Piazzale A. Moro 2, 00185 Roma, Italy.}}
\footnotetext{\textit{$^{b}$~Dipartimento di Fisica, "Sapienza" Universita' di Roma, Piazzale A. Moro 2, 00185 Roma, Italy. }}



\section{Introduction}

The investigation of effective interactions in soft matter has a long history and provides the main theoretical framework to treat highly complex systems with many degrees of freedom experiencing several time and length scales\cite{likos2001effective}. 
Since the pioneering work of Asakura and Oosawa on depletion interactions, the addition of a non-interacting cosolute (defined in the most general way as a macromolecular additive in suspension, including polymers, micelles, nanoparticles, etc.) has 
become a fundamental tool to manipulate and control the phase behavior of colloidal suspensions\cite{asakura1958interaction, vrij1976polymers, poon1993experimental, mao1995depletion, dijkstra1999phase, dijkstra1999phaseJCP, buzzaccaro2007sticky, lu08nature, lekkerkerker2011colloids, mahynski2014stabilizing,ashton2015porous, rovigatti2015soft}.
 More recently, it has become clear that also the cosolute properties can be exploited, not only to induce simple depletion effects, but also to give rise to new types of effective interactions among colloidal particles.


The prototype of this situation is that of colloids interacting through the so-called ``critical Casimir forces''\cite{fisher1978cr}, occurring whenever colloidal particles are in solution with a  solvent \cite{hertlein2008direct, bonn2009direct, edison2015critical, Huchtpre2015, ansini2016} or with a cosolute\cite{buzzaccaro2010critical,gnan2012properties, gnan2012tuning} close to a second-order critical point.
Such forces are indeed arising because of the confinement of thermal critical fluctuations between the surfaces of the colloids; this generates a  long-ranged effective potential which has, for fluids belonging to the Ising universality class, an exponential tail whose range is controlled by the correlation length of the critical fluid\cite{fisher1978cr,gambassi2009critical,gnan2012properties}. The experimental confirmation of critical Casimir forces\cite{hertlein2008direct, paladugu2016nonadditivity} has opened a new research avenue for generating colloidal aggregation \cite{zvyagolskaya2011criticality,bonn2009direct, buzzaccaro2010critical, BuzzaccaroJCP2011, gnan2012properties} and stable colloidal phases \cite{faber2013controlling,dang2013temperature,edison2015critical} since the range of the potential can be reversely modified by simply  tuning the temperature close to the critical point. 
Casimir-like effects are very general, occurring whenever the fluctuations of an arbitrary order parameter are confined between two surfaces. Besides the famous quantum effects, classical analogues exploit 
fluctuation-induced phenomena and have been recently explored under thermal gradients\cite{kirkpatrick2015nonequilibrium}, in active baths\cite{ray2014casimir,ni2015tunable}, in granular materials\cite{cattuto2006fluctuation}, in driven diffusive systems \cite{Kardarprl2015} , in perturbed near-critical mixtures\cite{furukawa2013nonequilibrium} or close to capillary condensation \cite{maciolekPRE2001,OkamotoJCP2012}. 

In a recent work \cite{gnan2014casimir} we have shown that it is also possible to exploit the analogy between a second order critical point and the percolation transition to induce a Casimir-like effective force among colloids. In particular, by performing numerical calculations of the effective potential between two hard sphere (HS) colloids immersed in a sol of clusters close to its gelation point,  we have found that the confinement of the clusters-size fluctuations between the colloids surface gives rise to a long-range effective force, whose properties depend on the universal nature of percolation in analogy with the critical Casimir effect where the force is controlled by the universal properties of critical phenomena. In this context, we have also shown that the force can also be interpreted as a depletion force exerted by a highly polydisperse medium. Indeed the resulting effective potential between two colloids can be described, for sufficiently low density, via a generalized Asakura-Oosawa model taking into account the polydispersity associated to the cluster size distribution of the sol. Interestingly, we found that the range of the effective potential is controlled by the connectivity length of the sol which reflects the clustering properties of the medium; this means that, independently on the density, whenever the cluster size distribution is the same, the effective potential close to percolation always displays the same range. These striking results occur when colloids are suspended in a so-called ``chemical'' sol, i.e. a cosolute in which the clusters have infinite lifetime and therefore can be treated as particles interacting  via  excluded volume  among them and with the colloids in solution. 
Such situation is relevant for specific experimental conditions where irreversible clusters can be achieved either by the use of chemical reactions in the sol or when the lifetime of clusters is significantly longer than the experimental time scales (i.e. experiments that probe only a single sol microstate). However often  clusters are reversible, breaking and reforming during the course of an experiment, and thus it is legitimate to ask how the results described above depend on the lifetime of the clusters in the sol.  

To tackle this question, in this work we study the behavior of the effective potential between colloids immersed in a ``physical'' sol, where clusters are fully reversible, for a series of state points at different temperatures and densities close to percolation, but having identical cluster size distribution. 
We show that such different state points do have different structural properties which are reflected in the effective potential arising when two colloids are put in the solution.  Thus, we find that the range of the effective potential increases when the sol becomes more and more correlated, because it is controlled by the thermal correlation length  (rather than by the connectivity length) of the sol. Comparing with the irreversible case, we thus find a drastic difference in the behavior of the effective potential. To rationalize this difference, we also study the effect of the bond lifetime by introducing an artificial barrier in the sol interaction as previously used in \cite{zaccarelli2003activated,saika2004effect,parmar2015kinetic}, which does not alter the structural properties of the sol but only its dynamics. We find that, as long as the bond lifetime remains finite, albeit long, colloids will always experience the reversible effective potential, which remains always distinct, physically and conceptually, from the irreversible one. 

\section{Model and Methods}
We first investigate the thermodynamic and geometric properties of the cosolute, that is modelled through a pairwise anisotropic Kern-Frenkel \cite{kern2003fluid} three-patches (3P) potential. This amounts to consider particles as hard spheres of diameter $\sigma$  with three attractive sites, located on the equator. The interaction $V_{\alpha\beta}^{SW}(|\vec r_{ij}|)$ between two sites $\alpha$ and $\beta$ of two particles $i$ and $j$ at a center-to-center distance $\vec r_{ij}$ is a square-well of width $\delta=0.119$ and depth $\varepsilon=1$
\begin{equation}\label{eq:SW}
	V_{\alpha\beta}^{SW}(|\vec r_{ij}|)=\left\lbrace
 	\begin{array}{l}
 		\infty\quad \text{ if}\quad |\vec r_{ij}|<\sigma,\\
 		-\varepsilon \quad \text{ if}\quad \sigma\le |\vec r_{ij}|\le \sigma (1+\delta),\\	 
		 0  \quad \text{ otherwise}
  	\end{array}
  	\right.
\end{equation}
\noindent modulated by an angular function $G$ that depends on the orientation of the particles:
\begin{equation}\label{eq:Angular_fun}
	G(\hat r_{ij},\hat r_{i\alpha},\hat r_{j\beta})=\left\lbrace
 	\begin{array}{l}

	   1 \text{ if}\quad\left\lbrace
		\begin{array}{l}
  	    	\hat r_{ij}\cdot\hat r_{i\alpha}>\cos(\theta_{max}),\\
      		-\hat r_{ij}\cdot\hat r_{j\beta}>\cos(\theta_{max}),\\
      	\end{array}
       \right.\\

	   0  \quad \text{ otherwise}
  	\end{array}
  	\right.
\end{equation} 

\noindent where $\hat r_{i\alpha}$ and $\hat r_{i\beta}$  are unit vectors from the centre of particle $i(j)$ to the centre of the patch $\alpha(\beta)$ on the surface. In the expression above $\cos(\theta_{max})$ accounts for the volume available for bonding and its value is set to $\cos(\theta_{max})=0.894717$. The total interaction potential reads as:
 
 \begin{equation}\label{eq:3P_pot}
	V_{ij,\alpha\beta}=V_{\alpha\beta}^{SW}(|\vec r_{ij}|)G(\hat r_{ij},\hat r_{i\alpha},\hat r_{j\beta}).
\end{equation}.

Here $\sigma$ and $\varepsilon$ are chosen as units of length and energy. We thus perform Monte Carlo (MC) simulations in the canonical ensemble of $N$ particles at several distinct temperatures and packing fractions.

To evaluate the effective potential $V_{eff}$ we perform Monte Carlo simulations of two large HS colloids (of diameter ten times larger than that of the cosolute particles) in solution with  $3P$ particles for several state points.
$V_{eff}$ is evaluated by constraining the two colloids to move along the $x$-axis of a parallelepipedal box where the length of the $x$-edge  $L_x =70\sigma$ is  twice and half  the length of $L_y$ and $L_z$. This guarantees that, for the simulated state points, the surface-to-surface distance  $r$ between colloids is always larger than the distance at which $V_{eff}$ goes to zero.
In order to reduce the computational time, we use an umbrella sampling technique that allows us to run parallel simulations in which
the two colloids sample only a limited range $\Delta_i$ of distances. In each simulation run, we evaluate the probability of finding the two colloids at a given distance $P(r,\Delta_i)$. Since different runs are allowed to have a small overlap in the probed $\Delta_i$, we can obtain the total $P(r)$ by splicing together  consecutive windows. Finally we extract the effective potential as $\beta V_{eff}=-ln(P(r))$, except for a constant that is set by imposing $V_{eff}(\infty)=0$.

\begin{figure}[h]
\begin{center}
\includegraphics[width=6cm]{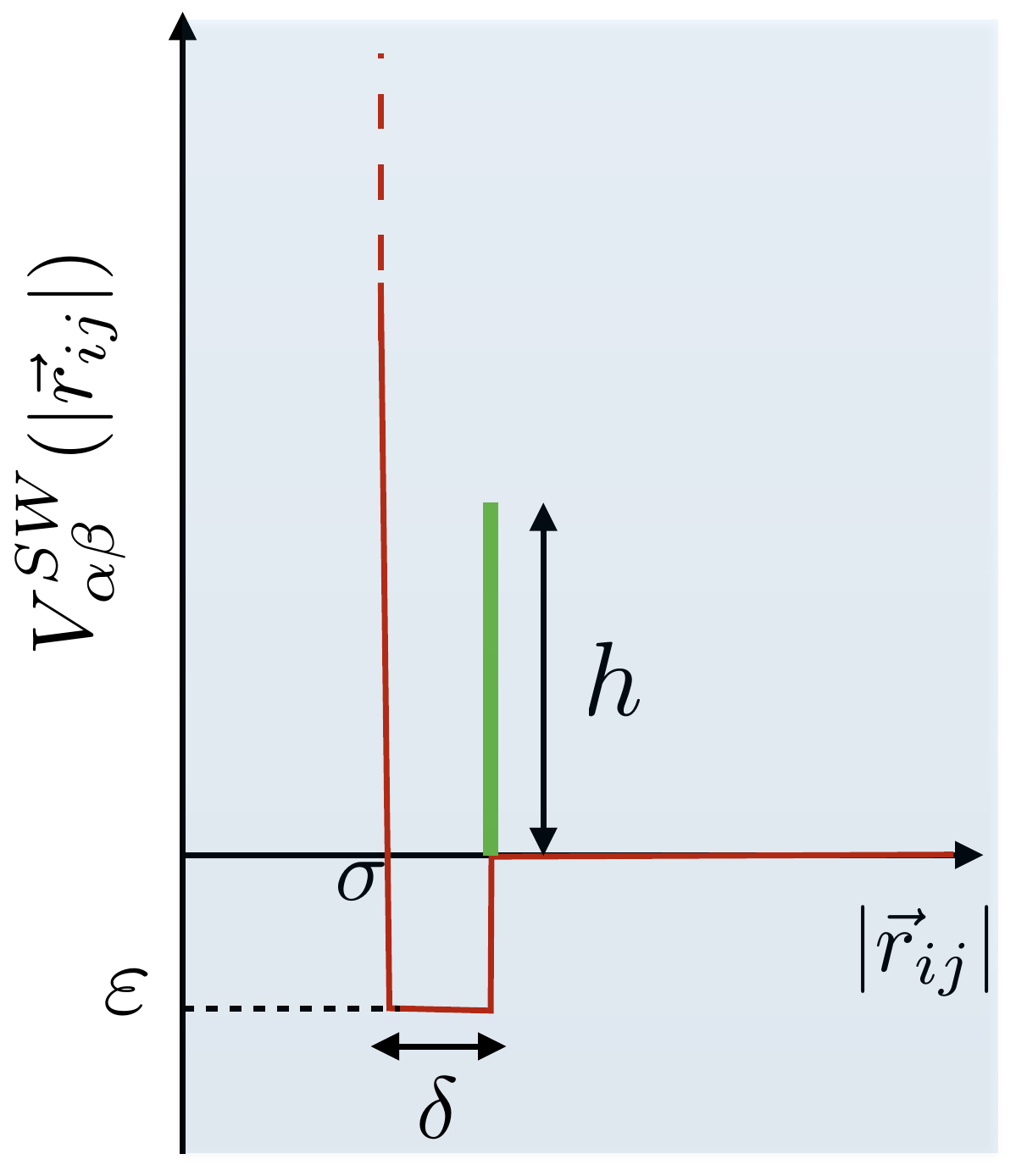}
\caption{Sketch of the modified SW interaction between 3P particles where an infinitesimally-thin barrier (green thick line) of finite height $h$ is also considered.}
\label{fig:barriersketch}
\end{center}
\end{figure}

Following \cite{saika2004effect}, to modify the bond lifetime without altering the thermodynamics, we add an infinitesimally thin barrier to the edge of the square well potential among patches as shown in Fig.\ref{fig:barriersketch}.
\noindent  The barrier has a finite height $h$ modulating the lifetime of the bonds among particles. We thus perform Monte Carlo simulations in the presence of the barrier by applying a modified Metropolis rule that account for the energy cost
 to overcome the barrier in forming /breaking bonds. \footnote{To account for the presence of the barrier we modify the Metropolis acceptance rule by introducing a penalty in the energy whenever bonds among particles are formed or broken. In particular, given the energy of the new configuration that needs to be accepted or refused, we account for broken bonds $N_{br}$ by adding a positive contribution $V_{br}=N_{br}\cdot(\varepsilon+h)$. Analogously, if new bonds $N_f$ are formed, the energy is modified  by the quantity $V_f=N_f\cdot h$. Note that the modified energies are only used in the Boltzmann factor entering in the Metropolis rule and do not correspond to real energies of the system.}

To measure the bond lifetime, we compute the bond autocorrelation functions, defined as
\begin{equation}
\phi_b(t)=\frac{\langle \sum_{i,j} n_{ij}(0) n_{ij}(t) \rangle}{N_b(0)} ,
\end{equation}
where $n_{ij}$ is 1 if particles $i,j$ are bonded, zero otherwise and $N_b(t)$ is the total number of bonds in the system at time $t$. Thus, the bond lifetime $\tau_b$ is defined as $\phi_b(\tau_b)=1/e$.

\section{Results}
\subsection{Properties of the cosolute along constant bond probability loci (iso-$p_b$ lines) }
\begin{figure}[h]
\begin{center}
\includegraphics[width=9.0cm]{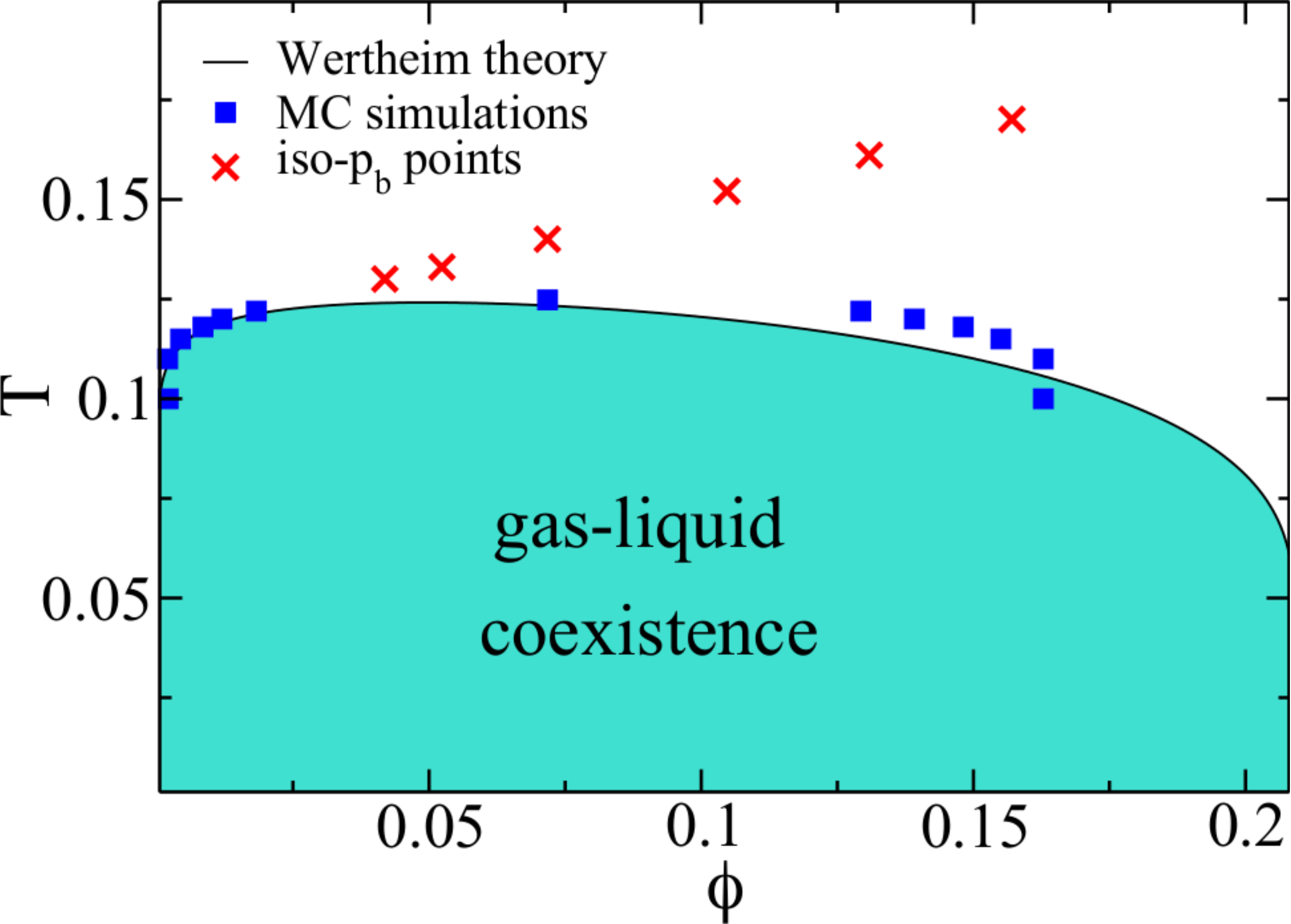}
\caption{Gas-liquid coexistence line (blue squares) of 3P particles evaluated from Monte Carlo simulations\cite{gnan2012surface} and theoretical calculations (black line) using Wertheim theory. The critical point is found at $T=0.125$, $\phi=0.072$. The figure also shows the simulated state points (red crosses) of constant bond probability $p_b$ with $p_b \sim 0.565$ investigated in this work. 
}  
\label{fig:percolation}
\end{center}
\end{figure}

 The $3P$ system belongs to the class of low-valence patchy models\cite{bianchi2006phase}, thus displaying a gas-liquid critical point at low densities and the possibility to form equilibrium gels. The phase diagram of the 3P model is shown in Fig.~\ref{fig:percolation} together with the simulated state points of this work. The gas-liquid coexistence line evaluated from Wertheim theory \cite{wertheim1984fluidsI,wertheim1984fluidsII,wertheim1986fluidsIII, wertheim1986fluidsIV} is reported together with the corresponding simulation results\cite{gnan2012surface}.

\begin{figure}[h]
\begin{center}
\includegraphics[width=9.0cm]{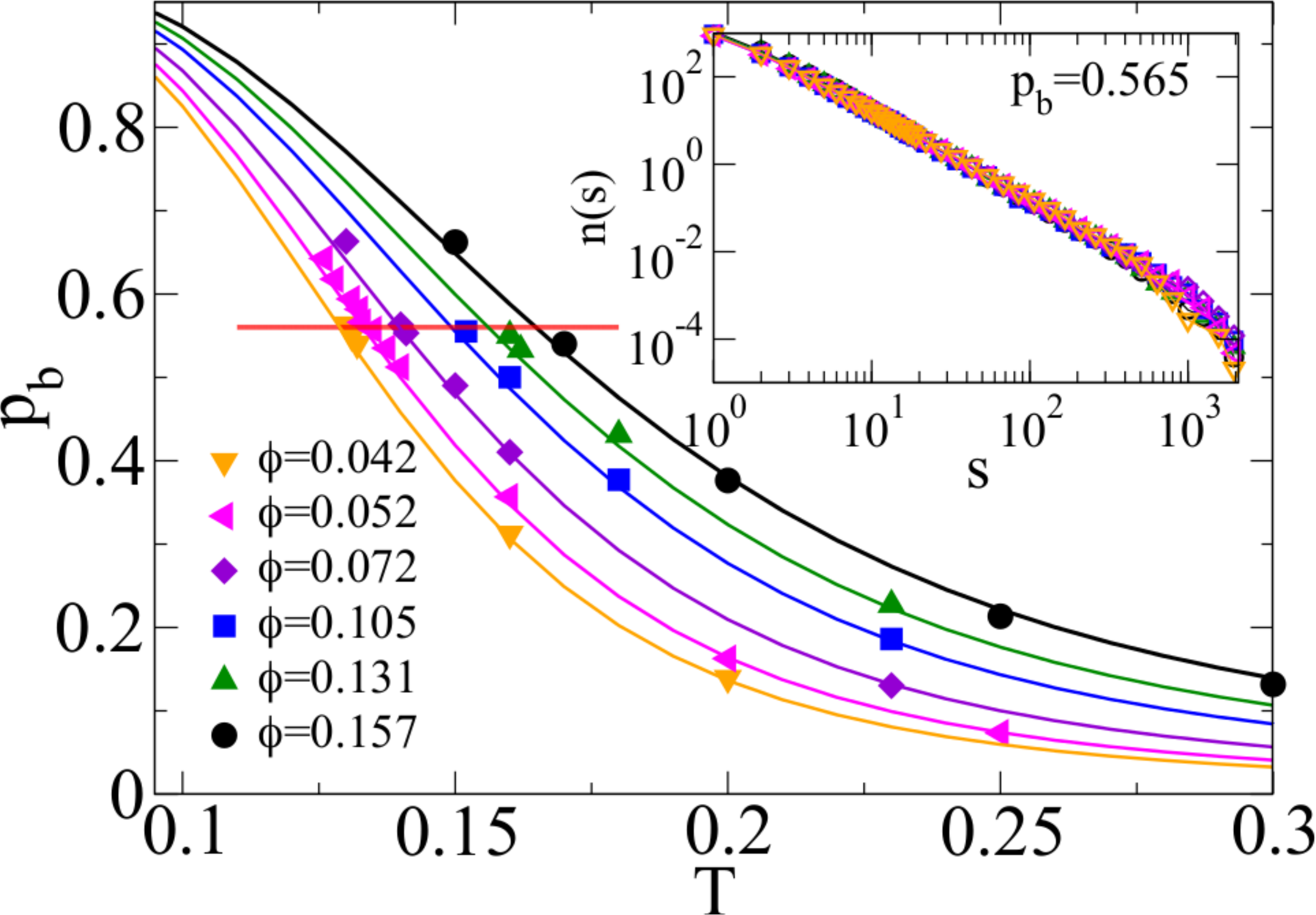}
\caption{Bonding probability $p_b$ as a function of T for different $\phi$ values. Symbols are results from MC simulations while solid lines are predictions from Wertheim theory. The dashed lines selects state points with $p_b\simeq 0.565$. Inset: Cluster size distribution $n(s)$  for different $\phi$ along the $p_b \sim 0.565$  locus. $n(s)$ is the same within the numerical precision.}
\label{fig:pbAnalysis}
\end{center}
\end{figure}
\begin{figure}[h!]
\centerline{\includegraphics[width=9cm]{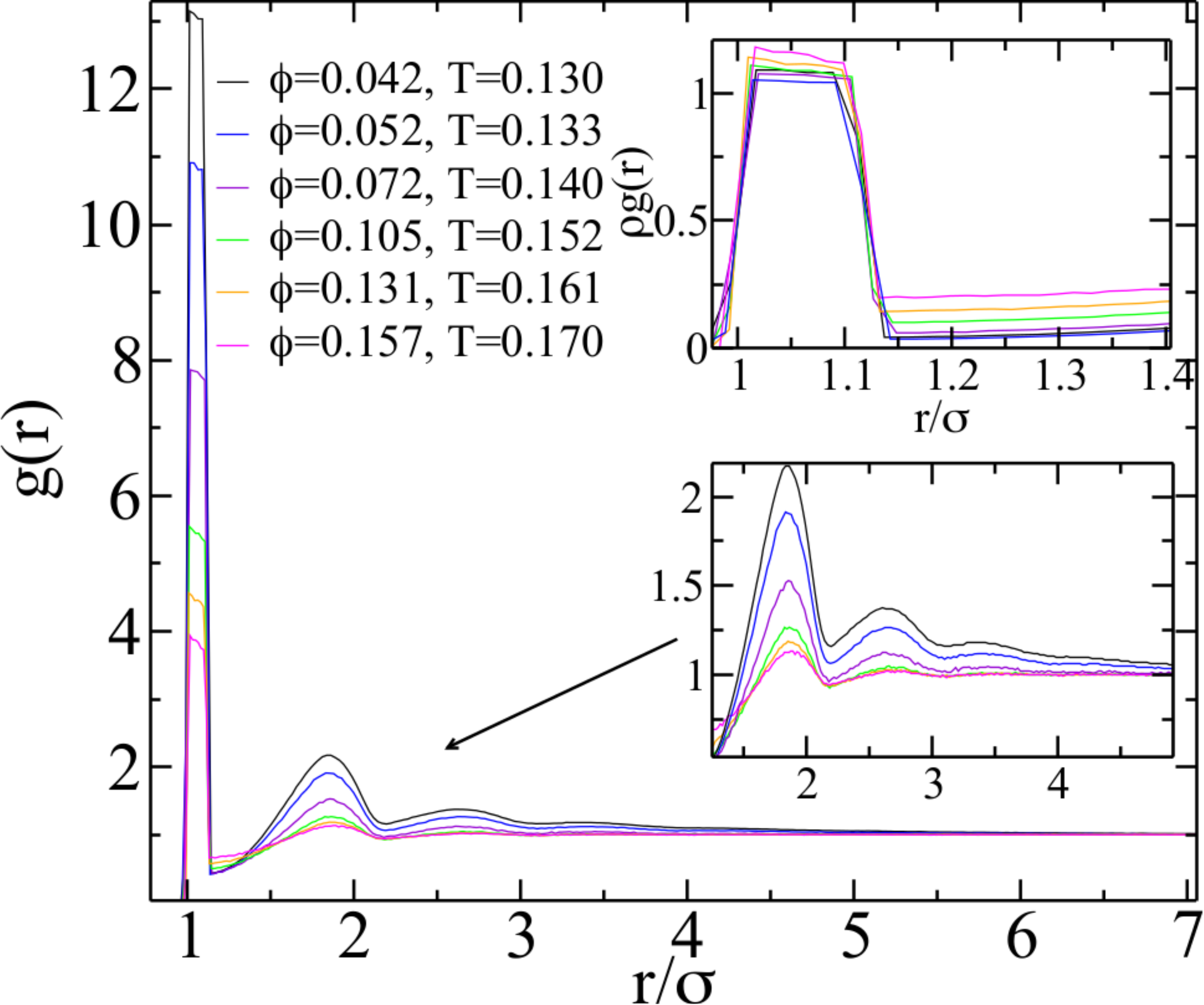}}
  \caption{Radial distribution functions $g(r)$ along the iso-$p_b$ line. Insets: (upper panel) The iso-$p_b$ condition is  highlighted by scaling $g(r)$ for the number density $\rho$; (lower panel) magnification at longer-$r$ of the $g(r)$ along the iso-$p_b$ line.}\label{fig:gr}
\end{figure}
\begin{figure}[h!]
\centerline{
\includegraphics[width=9cm]{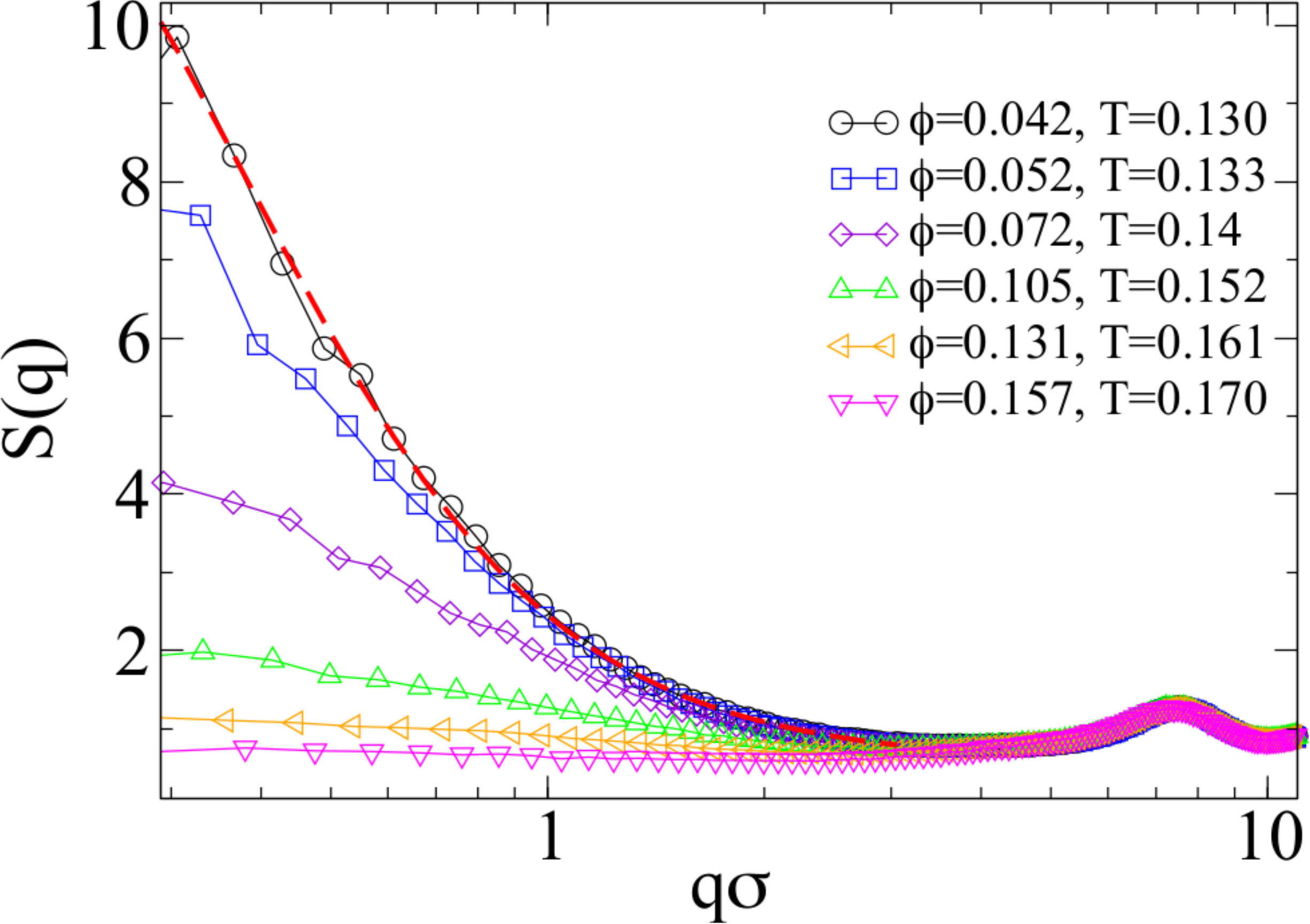}
}
\caption{Static structure factors $S(q)$ as a function of wavevector $q$ along the iso-$p_b$ line with $p_b \sim 0.565$. Fitting the low $q$ behavior to a Lorentzian\cite{hansen1990theory} (dashed line shown in the figure as an example), we extract the thermal correlation length $\xi$ for each state point.}\label{fig:sq}
\end{figure}

\noindent 
 Since the connectivity properties of the sol (e.g. the probability of forming bonds) depend on temperature $T$ and  packing fraction $\phi=\frac{\pi}{6} \rho \sigma^3$ (where $\rho$ is the number density). It is possible to tune $T$ and $\phi$ in order to find state points of equal bond probability ($p_b$), drawing a line in the gas-liquid phase diagram which can be considered a precursor of the percolation line \cite{bianchiJPCB2007, bianchi2008equilibrium}. 
Figure ~\ref{fig:pbAnalysis} shows $p_b$ as a function of $T$ for different densities. Solid lines are predictions from Wertheim's theory which well interpolate numerical results for a wide range of densities as shown also in \cite{gnan2012surface}. For  the model employed here, the bond probability is given by $-E/(f N)$ where $E$ is the total potential energy, $N$ is the total number of particles and $f$ is the maximum number of bonds that a particle can form: for a $3P$ particle $f=3$. 
Note that the percolation transition for this model has been estimated to take place at $p_c\simeq 0.618$ \cite{gnan2014casimir}.
We thus select state points having a bond probability $p_b=0.565\pm0.010$ at a relative distance to percolation $p_b/p_c = 0.92$, which are also reported  in Fig.~\ref{fig:percolation}. Along this line the distribution $n(s)$ of clusters of size $s$ is found to be the same within numerical precision (inset of Fig. ~\ref{fig:pbAnalysis}) in agreement with Flory-Stockmayer predictions\cite{sciortino2011reversible}, valid in the absence of  bond loops, stating that the cluster size distribution only depends on $p_b$. This means that upon varying $T$ and $\phi$ along the chosen iso-$p_b$ line, particles aggregate into clusters of similar polydispersity  with the same probability.
Although the aggregation properties are similar, the structure of the sol is quite different along the iso-$p_b$ line as shown by the radial distribution functions $g(r)$ and by the static structure factors $S(q)$
reported respectively in Fig.~\ref{fig:gr} and ~\ref{fig:sq}.

\noindent The different structure of the sol reflects the progressive dilution of the clusters and 
 the increased proximity to the gas-liquid spinodal at low $\phi$ and $T$. Both effects contribute to the small wave vector increase of $S(q)$. 
Similarly the  radial distribution functions $g(r)$ becomes progressively more structured on lowering $\phi$, as shown  in Fig.~\ref{fig:gr} (main panel and lower inset). 
Two features  are interesting: 
the intense peak between $\sigma$ and $\sigma(1+\delta)$  reflecting the presence of  bonded pairs and the onset of a long range correlation, highlighted by the approach of
the radial distribution function to one from above.   These two features are indicated in more details in the two insets.
In the upper inset of Fig.~\ref{fig:gr} all the peaks of $\rho g(r)$ superimpose to a good approximation consistently with the iso-$p_b$ constraint. 

From the static structure factors we obtain an accurate estimate of the thermal correlation length $\xi$ through a fit of the low wavevector region with a Lorentzian  function~\cite{gnan2012properties}, as shown in Fig. ~\ref{fig:sq}.
It is worth to stress that the proximity of the critical point to the iso-$p_b$ line might influence the dynamics of the low-$T$ state points. In a previous work \cite{gnan2012properties}, we have estimated that, at the critical packing fraction, state points belong to the critical region when $T\leq1.045\,T_c\sim0.131$. Hence  only the point at the lowest investigated $\phi$ could be influenced by critical fluctuations. This is supported by the value of $\xi\sim2.4$ extracted from the $S(q)$  which is consistent with previous results at a similar temperature but at the critical packing fraction (rather than along an iso-$p_b$ line). 



\subsection{Effective potentials along iso-$p_b$ lines for reversible clusters}
\begin{figure}[h!]
\begin{center}
\includegraphics[width=9cm]{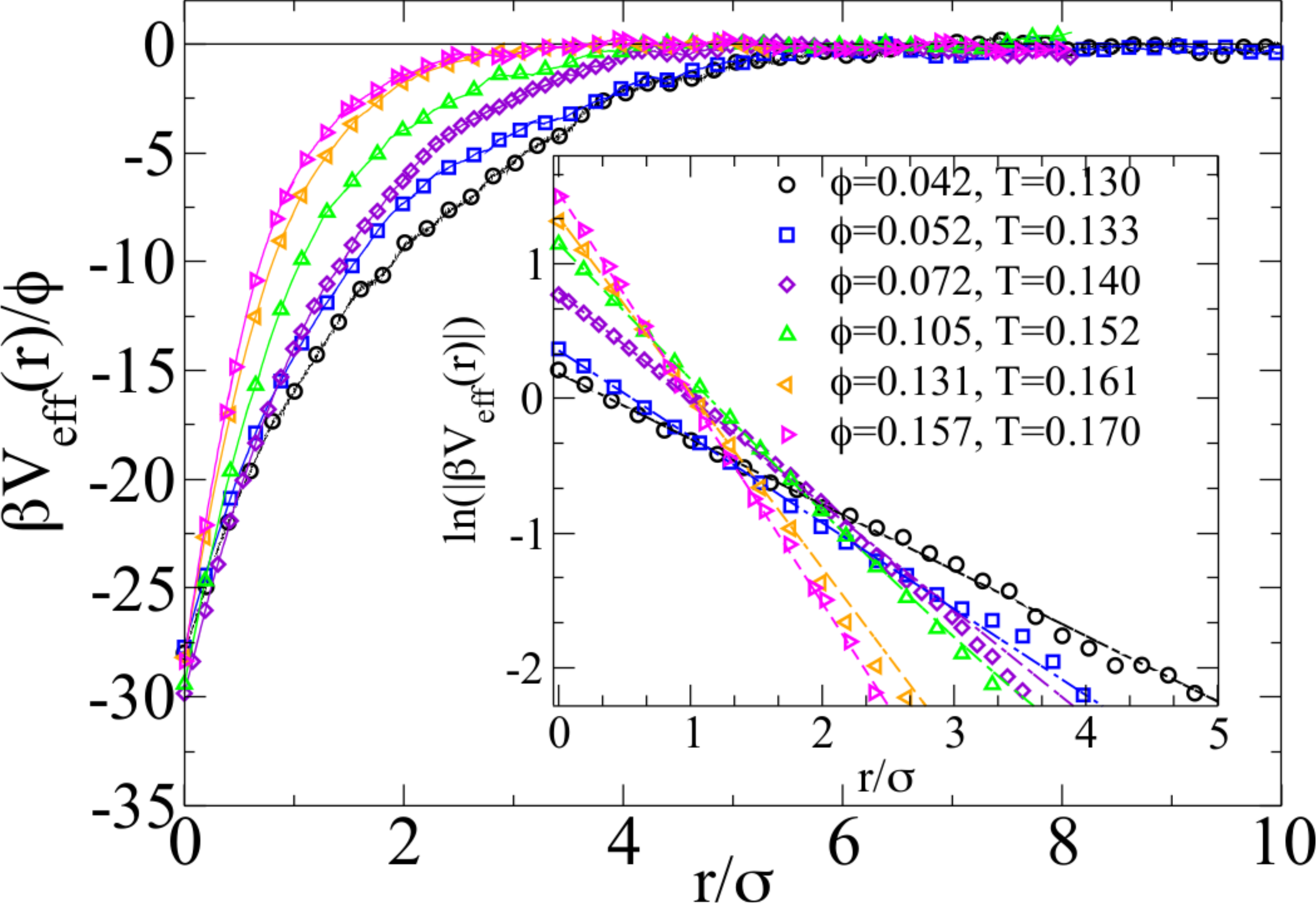}
\caption{ Effective potentials for six state points along the iso-$p_b$ line with $p_b \sim 0.565$. Potentials are scaled by $\phi$ to show that close to contact they are controlled by the density of depletants as in standard depletion. The inset shows the logarithm of the effective potentials and the corresponding linear fits (dashed lines). The slopes of the fits correspond to the inverse of the range of the potentials.}
\label{fig:rev_pots}
\end{center}
\end{figure}

The effective potentials $V_{eff}(r)$ of two HS colloids immersed in the 3P sol along the chosen iso-$p_b$ line are shown in Fig.~\ref{fig:rev_pots}.
First of all, we observe that  the value of the effective potentials at contact  ($V_{eff} (r=0)$) scales with 
with the sol packing fraction, as usually found in standard depletion. Interestingly, the range of the potentials is larger than the cosolute particle size $\sigma$ which sets the lenght scale of $V_{eff}$ in  low-$\phi$ depletion phenomena by non-aggregating co-solutes. However, we notice that, moving along the iso-$p_b$ line, the range of $V_{eff}$ increases by decreasing the cosolute $\phi$ and $T$ (as magnified in the inset of Fig.~\ref{fig:rev_pots}). Thus, although particles organize into the same distribution of clusters, the contributions of these clusters to the effective potentials are different depending on the considered state point.

\begin{figure}[h!]
\centerline{
\includegraphics[width=9cm]{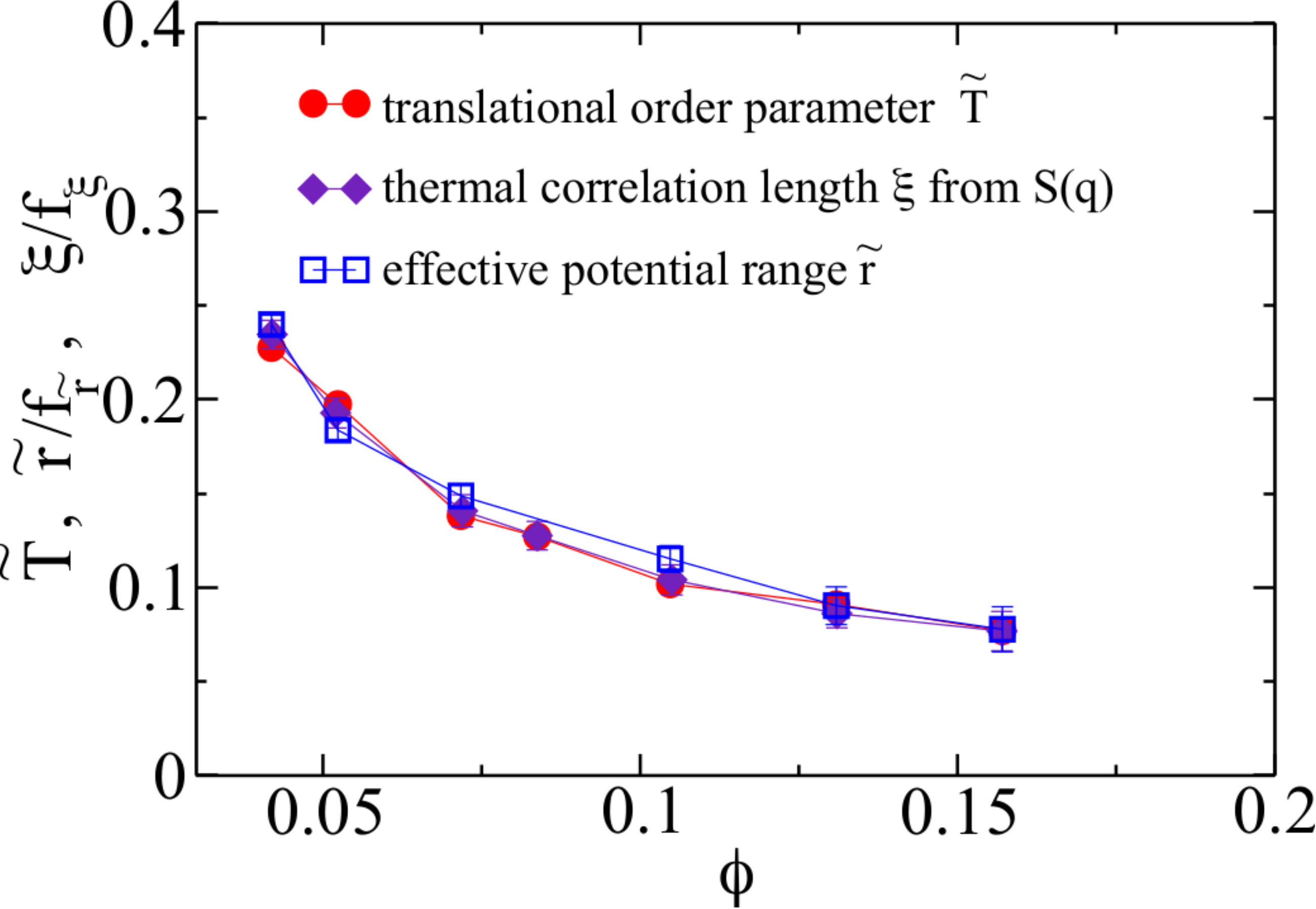}
}
\caption{Translational order parameter $\tilde{T}$, thermal correlation length $\xi$ and range of reversible effective potential $\tilde{r}$  as a function of $\phi$  along the chosen iso-$p_b$ line. Both $\xi$ and $\tilde{r}$ have been scaled by a factor of, respectively, $f_{\xi}=11.33 \sigma$ and $f_{\tilde{r}}=9.37 \sigma$ to superimpose them with $\tilde{T}$. Error bars are estimated by the fits of $S(q)$ at low $q$ for $\xi$ and of $V_{eff}$ for $\tilde{r}$. }
\label{fig:range}
\end{figure}

\noindent In the following, we try to connect the observation on the increasing range of $V_{eff}$ with the behaviour of the cosolute along the iso-$p_b$ line. To do so, we have evaluated the so-called translational order parameter $\tilde{T}$ (not to be confused with temperature), following Ref.~\cite{truskett2000towards}.  This is defined as
\begin{equation}
\label{eq:transl}
\tilde{T}= \frac{\int_{\sigma \rho^{1/3}}^{r_c} |g(r')-1| dr'}{r_c-\sigma \rho^{1/3}}
\end{equation}
where $\rho$ is the number density, $r'=r\rho^{1/3}$ and $r_c$ is a cutoff value. We have chosen $r_c=5\sigma$.
 This observable provides a measure of the thermal correlation length, as shown in Fig.~\ref{fig:range} where the behavior of $\tilde{T}$ is reported
together with $\xi$ obtained from the static structure factors, scaled  to superimpose it with  $\tilde{T}$. Both observables display an identical behavior, showing a clear increase with decreasing $\phi$ (and $T$) along the chosen iso-$p_b$ line due to the increasing proximity to the gas-liquid spinodal. 
To rationalise the fact that the observed range increases with decreasing $\phi$, we have estimated the range $\tilde{r}$ of the effective potentials by performing exponential fits, highlighted in the inset of Fig.~\ref{fig:rev_pots}. Thus, in Fig.~\ref{fig:range} we also compare the behavior of $\tilde{r}$ with $\tilde{T}$ upon moving along the iso-$p_b$ line finding  again 
that the two parameters have the same $\phi$ dependence. The agreement between the two different observables suggests that  the growth in the correlations among the sol particles expressed in the structural correlators is also responsible of the growth of the range of the effective potential. 
This picture holds for all studied state points, including the lowest $\phi$, which is partially influenced by critical fluctuations.
These results allow us to conclude that, in the case of reversible bonds, the effective potential is controlled by thermodynamic, rather than connectivity, properties. 

\noindent 

\begin{figure}[h!]
\begin{center}
\includegraphics[width=9cm]{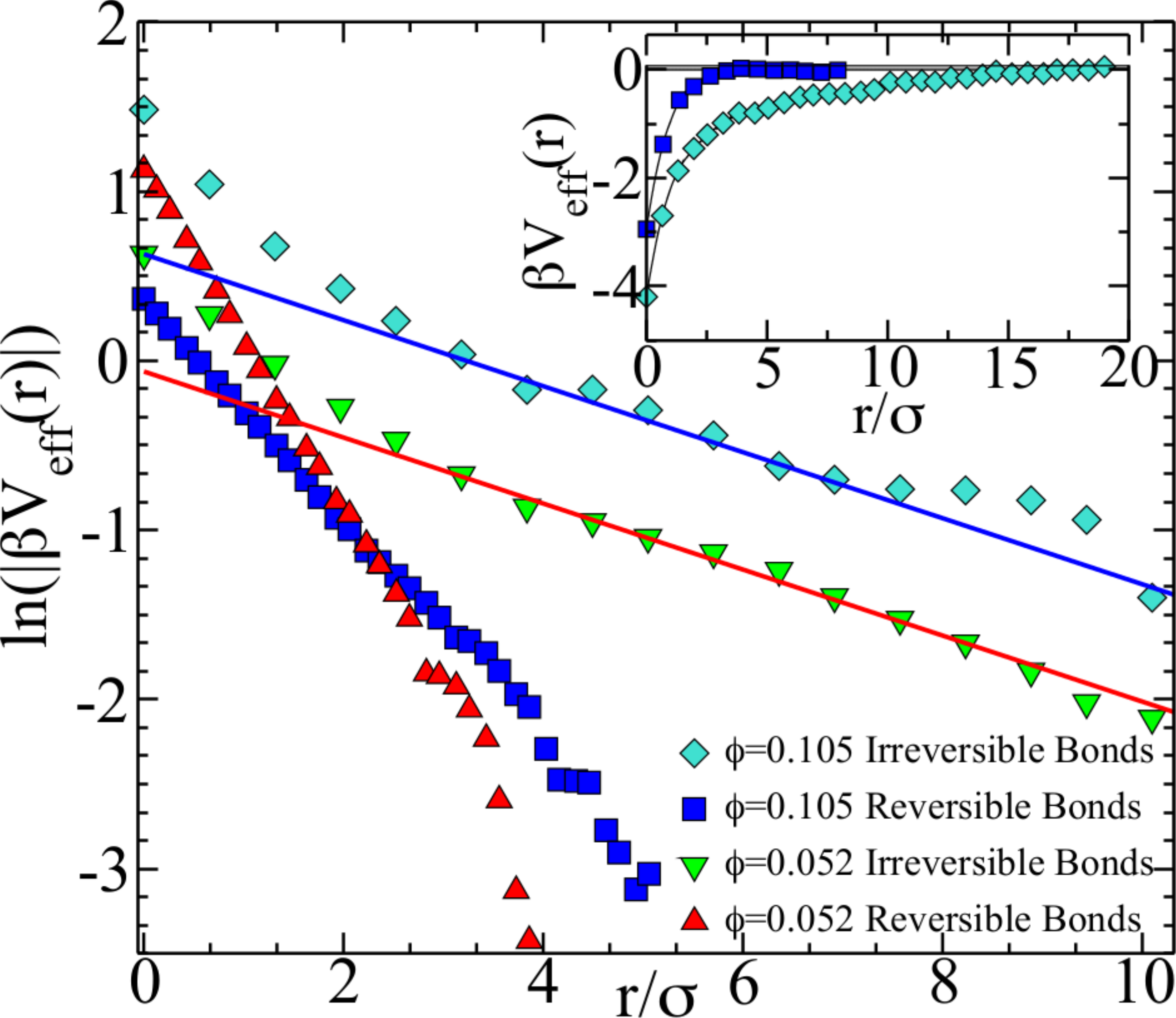}
\caption{Effective potentials for reversible and irreversible clusters for two different state points along the chosen iso-$p_b$ line. Straight lines are exponential fits to the tails of the irreversible $V_{eff}$, having the same decay length for both state points. Inset: comparison of $V_{eff}$ for $\phi=0.105$ for reversible and irreversible clusters.}
\label{fig:irrev_range}
\end{center}
\end{figure}

We recall that, if the cluster bonds have infinite life time, the scenario is significantly different, since in that case  is the connectivity length of the clusters in the sol which controls the range of the effective potentials~\cite{gnan2014casimir}. 
Thus, when two colloids are immersed in such sol of clusters they experience an effective potential whose range is independently from $\phi$ but depends only on the connectivity properties of the system. This is shown in Fig.~\ref{fig:irrev_range}, where the effective potentials in the case of irreversible clusters are shown in log scale for two state points along the iso-$p_b$ line together with the corresponding potentials for the reversible clusters case. 
We find that the decay of the two potentials for irreversible clusters are well described by exponential functions having the same decay length as shown in the figure, highlighting that the range is the same along the iso-$p_b$ lines. This is of course different from the behaviour for the reversible cluster case. The inset of Fig.~\ref{fig:irrev_range} also shows that the range of $V_{eff}$ for irreversible clusters is by far larger than that for reversible ones. 
This difference between reversible and irreversible clusters already indicates that for reversible clusters the
 observed effective potentials cannot be modeled in terms of ideal depletion from a polydisperse sol, despite we have found that
$\beta V_{eff}(0)$ scales with the sol density, as predicted by standard depletion approaches. 
Finally we note that especially at low $\phi$, the range of $V_{eff}$ is larger than $2\sigma$, thus three-body effects may play an important role in the colloid-colloid effective interaction. Recent studies have focused on the role of three-body interactions in critical Casimir forces \cite{paladugu2016nonadditivity,MattosJCP2013}; we expect a similar role in the present case. Still we note that the ratio between the colloid size and the effective potential range can be controlled either by tuning the colloid diameter or by tuning the average cluster size distribution (hence $p_b$). When this ratio is smaller than about $10\%$  we expect that the three-body interactions will become less important, in analogy with results derived for standard depletion \cite{dijkstra1999phase}.

\subsection{Effect of the bond lifetime on the effective potentials}
So far, we have established that colloids immersed in a sol forming respectively reversible or irreversible clusters experience rather different effective potentials. In the first case, the range of the effective potential is set by the thermal correlation among particles and seems not to be mediated by clusters, while in the second case clusters act as depletants and indeed the resulting $V_{eff}$ has been successfully interpreted through a depletion model. Thus, one could be tempted to ask whether the lifetime of clusters plays any role, and if such clusters become sufficiently long-lived, whether they contribute to the effective potentials, for example providing a different $V_{eff}$ bridging the fully reversible (thermal) and irreversible (geometric) case in a continuous way.

\begin{figure}[h!]
\begin{center}
\includegraphics[width=9cm]{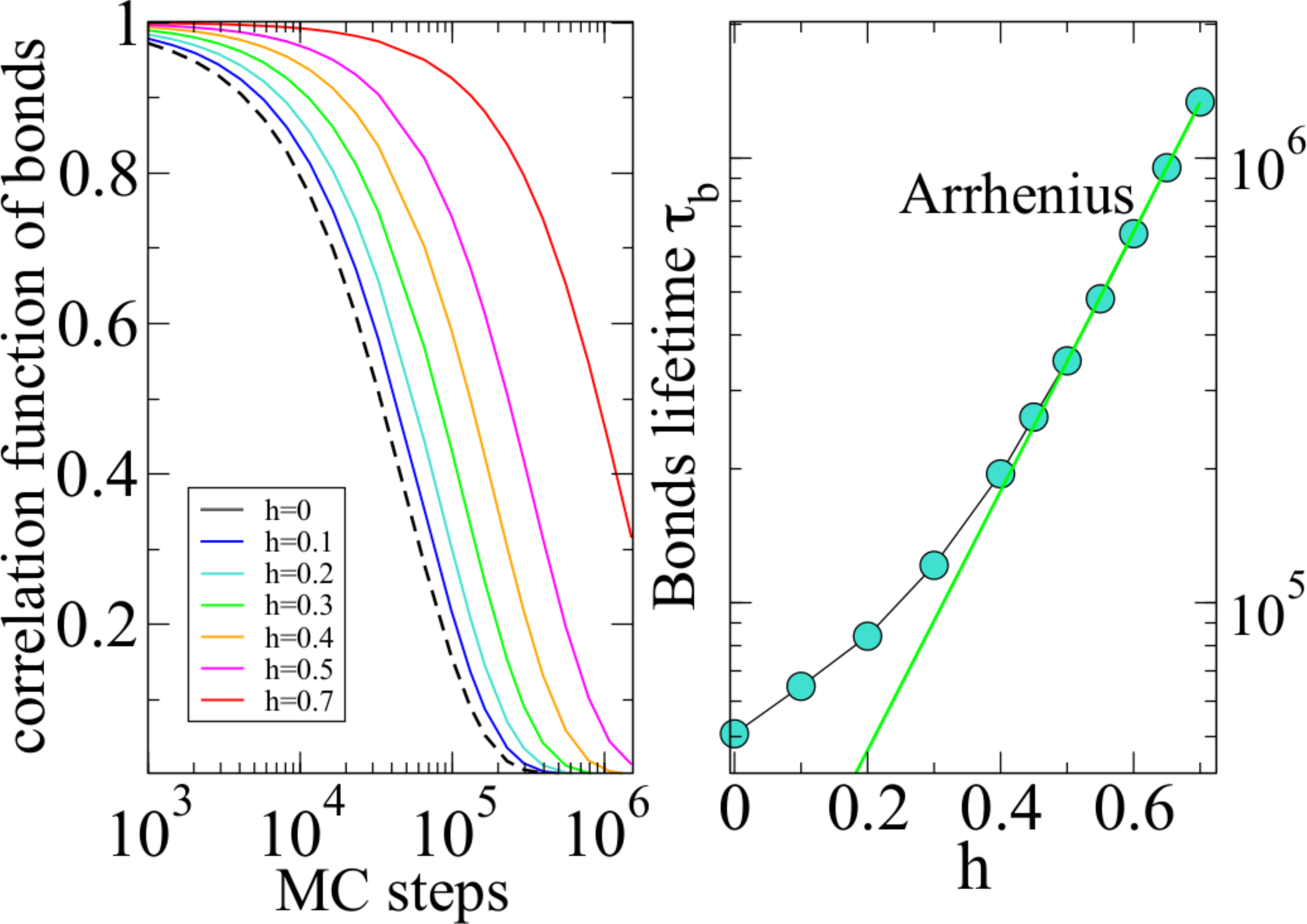}
\caption{Left panel: bond correlation function for a 3P sol in the presence of an additional barrier with several values of height $h$ for $\phi=0.052$ and $T=0.133$. Right panel: the associated bond lifetime $\tau_b$ as a function of $h$, highlighting the onset of Arrhenius behavior at large enough (i.e.  $h > 0.5$) barrier heights. }
\label{fig:bondsbarrier}
\end{center}
\end{figure}

To this aim, we have added an infinitesimal barrier to the 3P square-well interaction~\cite{saika2004effect}. First, we have evaluated the properties of the system for several values of
the height $h$ of the barrier.  While we find that structural and thermodynamic properties, such as $g(r)$ and energy, are identical upon increasing $h$, the MC  dynamics of the system is largely affected by the introduction of the barrier, in agreement with previous works\cite{zaccarelli2003activated,saika2004effect,parmar2015kinetic}. This is shown in Fig. 
~\ref{fig:bondsbarrier} where the bond correlation function is reported for several $h$ values. The associated bond relaxation time $\tau_b (h)$ is also reported, displaying a clear crossover to an Arrhenius behavior at large enough $h$.

\begin{figure}[h!]
\begin{center}
\includegraphics[width=9cm]{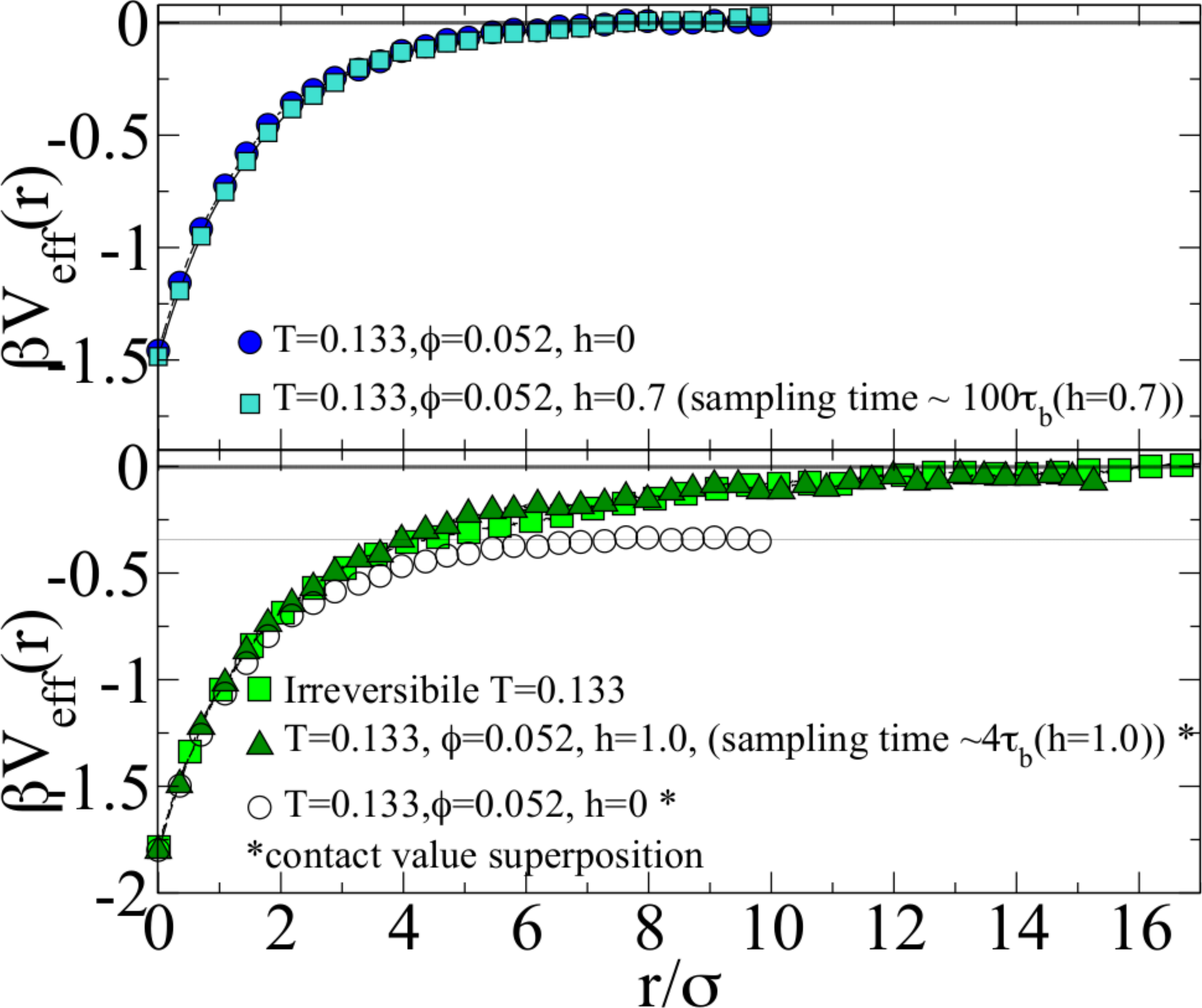}
\caption{(Top panel) Effective potential for two colloids immersed in a 3P sol at $\phi=0.052, T=0.133$  for reversible clusters with ($h=0.7$) and without ($h=0$) the addition of an infinitesimal barrier, which modulates the bond (and clusters) lifetime. The effective potential of reversible clusters with the barrier has been evaluated along a  time hundred times larger than the bond life time of the clusters.
(Bottom Panel) Effective potential for two colloids immersed in a 3P sol at $\phi=0.052, T=0.133$  for irreversible clusters and for reversible clusters with ($h=1.0$)  and without ($h=0$) the addition of an infinitesimal barrier. $V_{eff}(h=1.0)$ has been evaluated along a time of the order of the bond life time of the clusters. To show that  $V_{eff}(h=1.0)$ is inherently different from the case in which the potential is sampled for times much larger than the bond relaxation time, we compare it with $V_{eff}(h=0)$ by superimposing the their contact values, i.e. the points less affected by statistical error.}
\label{fig:potbarrier}
\end{center}
\end{figure}

We thus repeat the evaluation of the effective potential in the presence of the barrier, for $h=0.7$, where   $\tau_b$   is more than one order of magnitude larger with respect to the fully reversible case ($h=0$). The resulting $V_{eff}$ is reported in Fig.~\ref{fig:potbarrier}(top panel) together with the $h=0$ case. We find that the two potentials are identical, i.e. no effect of the bond lifetime of clusters is recorded on the effective potentials. Thus, $V_{eff}$ appears to jump discontinuously from the behaviour for reversible clusters  to the behavior for irreversible clusters.

However we expect that, if the bond lifetime is significantly larger (even if not infinite) than the time of the sampling, the system will still carry a 'memory' of the frozen cluster size distribution. To clarify this point we have performed MC simulations where the barrier height is $h = 1.0$. Under this condition, the bond life time is so large (one order of magnitude larger than $\tau_b(h=0.7)$) that we are not able to follow the cluster break-up and re-bonding toward complete equilibration. We have thus measured $V_{eff}$ for $h = 1.0$ over a narrow time interval, which is long-enough to ensure a reduced noise in the sampling, but too short to observe complete cluster reorganisation. We have chosen this time to be $4 \,\tau_b(h=1.0)$ where $\tau_b \simeq 1\cdot 10^7$ MC steps. 

The result displayed in Fig.~\ref{fig:potbarrier}(bottom panel)  shows that,  when the sampling of the effective forces is performed over a time window smaller than the full relaxation of the system (i.e.  off-equilibrium), $V_{eff}$  coincides with the irreversible result.  This confirms that on time scales in which full equilibration of the cosolute is not achieved, the same results as for the chemical gel case are recovered.   


\section{Conclusions}
In this work we have investigated the effective potentials among colloids in a self-assembling cosolute. In particular, we have studied state points where the cosolute is close to percolation, moving along a path where the bond probability is constant and the cluster size distribution is identical. The resulting effective potentials are found to depend, at short distances, on the density of the cosolute, in agreement with depletion arguments. But at long distances they display an increasing range of interaction upon decreasing $\phi$ and $T$. 
This result bears some analogies with the Critical-Casimir effect: close to the critical point, in fact a critical fluid confined between two colloids gives rise to an effective interaction that can be theoretically described in terms of a universal exponential function whose decay is controlled by the critical correlation length. Far from the critical point the physical effect is the same but the effective interaction depends very much on the cosolute properties and it is not possible to derive a functional form that describes the behaviour of the effective interaction. Along the studied  iso-$p_b$ line,  the association of the cosolute is strong enough to give rise to completely attractive effective potentials and thus the exponential function seems to be a reasonable approximation for extracting a correlation length in analogy with critical Casimir forces.  However, at higher temperatures, a fit with an exponential function would be less and less accurate until, at very-high $T$ (where co-solute particles behave as hard-spheres), the resulting effective potential would show oscillations as in mixtures of highly asymmetric HS particles.

The reversible clusters results differ with previous findings for irreversible clusters (despite the same cluster size distribution and average density), which did not  show any dependence in the range on the chosen state points. This is because under conditions of irreversible bonds the potentials are slaved to the geometric properties of the clusters, which act as non-interacting highly polydisperse rigid bodies generating a depletion-like potential on the colloids.
Our finding show that when clusters can break and reform on timescale shorter than the experimental time scale, the cluster identity is lost 
and  the effective potentials depends only on the thermal correlation length  of the sol. 

In order to identify whether the case of irreversible bonds represents a singular limit or it can be accessed in a continuous way, we have also investigated the effect of bond lifetime, which we appropriately tune by introducing an infinitesimal barrier in the cosolute attractive potential. We find that, even in the case of a large but finite lifetime, the resulting effective potentials experienced by colloids coincide with those obtained for reversible clusters where the bond lifetime is short. Thus, the bond lifetime, whenever finite, does not play any role on $V_{eff}$,  as long as the system 
can probe all micro states  of the sol, e.g. as far as  the measurement time of $V_{eff}$ is long enough. Under these conditions, colloids experience the same effective interactions independently of how often cluster break and reform. In this respect, the irreversible case remains a distinct physical situation, and one cannot interpolate continuously from one case to another.
Oppositely, if the observation time is smaller or comparable to the bond-life time (e.g. in out of equilibrium conditions), the colloids effective interaction will carry a memory of the initial cluster size distribution, and it will be similar to the irreversible interaction.

Our work has a direct relevance for experimental situations when aggregation of a second component, being it a solvent or a cosolute, is exploited to induce ad-hoc effective interactions. Only when the sol is chemical, forming clusters that cannot be broken, a Casimir-like effect induced by cluster confinement is found. Such situation can be achieved in systems of hyper-branched polymers or in which the number of bonds in the system can be controlled by irradiation. For physical sols, a shorter-ranged potential which is dictated solely by thermal properties of particles (and not of clusters) is recovered. Thus, in most physical realizations, the ``physical'' picture applies, which is not simply interpreted in terms of ideal depletion arguments, except for short colloid-colloid distances. 

\section{acknowledgements}
NG and EZ acknowledge support from MIUR (``Futuro in Ricerca'' ANISOFT/RBFR125H0M) and from the European Research Council
(ERC Consolidator Grant 681597, MIMIC).  FS and EZ acknowledge support from ETN-COLLDENSE (H2020-MCSA-ITN-2014, Grant No. 642774).

\balance


\bibliography{ReversibleBib} 
\bibliographystyle{rsc} 

\end{document}